\documentclass[letterpaper, 10 pt, conference]{ieeeconf}  % Comment this line out if you need a4paper

\IEEEoverridecommandlockouts                              % This command is only needed if 
\title{\LARGE \bf
Enhancing Network Intrusion Detection Systems: A Multi-Layer Ensemble Approach to Mitigate Adversarial Attacks
}

\author{
\begin{tabular}{c}
Nasim Soltani$^{1,3}$, 
Shayan Nejadshamsi$^{4}$, 
Zakaria Abou El Houda$^{1,3}$, 
Raphaël Khoury$^{2,3}$,\\
Kelton A. P. Costa$^{5}$, 
Tiago H. Falk$^{1,3}$, 
Anderson R. Avila$^{1,3}$
\end{tabular} \\[1ex]
\small
$^1$Institut national de la recherche scientifique (INRS-EMT), Université du Québec, Montréal, Québec, Canada\\
$^2$Department of Computer Science and Engineering, Université du Québec en Outaouais, Gatineau, Québec, Canada\\
$^3$INRS-UQO Mixed Research Unit on Cybersecurity, Gatineau, Québec, Canada\\
$^4$Concordia Institute for Information Systems Engineering (CIISE), Concordia University, Montréal, Québec, Canada\\
$^5$Department of Computing, São Paulo State University, São Paulo, Brazil
}

\usepackage{graphicx}
\usepackage{xcolor}
\usepackage{pifont}
\usepackage{amsmath}
\usepackage{array}
\usepackage{mdframed}
\usepackage{tcolorbox}
\usepackage{subcaption}
\usepackage{cite}
\usepackage{hyperref}

\newcommand{\cmark}{\textcolor{green!60!black}{\ding{51}}}
\newcommand{\xmark}{\textcolor{red}{\ding{55}}}

\newtcolorbox{mybox}
{colback=red!5!white,colframe=red!75!black}

\begin{document}

\maketitle

\thispagestyle{empty}
\pagestyle{empty}

%%%%%%%%%%%%%%%%%%%%%%%%%%%%%%%%%%%%%%%%%%%%%%%%%%%%%%%%%%%%%%%%%%%%%%%%%%%%%%%%
\begin{abstract}
Adversarial examples can represent a serious threat to machine learning (ML) algorithms. If used to manipulate the behaviour of ML-based Network Intrusion Detection Systems (NIDS), they can jeopardize network security. In this work, we aim to mitigate such risks by increasing the robustness of NIDS towards adversarial attacks. To that end, we explore two adversarial methods for generating malicious network traffic. The first method is based on Generative Adversarial Networks (GAN) and the second one is the Fast Gradient Sign Method (FGSM). The adversarial examples generated by these methods are then used to evaluate a novel multilayer defense mechanism, specifically designed to mitigate the vulnerability of ML-based NIDS. Our solution consists of one layer of stacking classifiers and a second layer based on an autoencoder. If the incoming network data are classified as benign by the first layer, the second layer is activated to ensure that the decision made by the stacking classifier is correct. We also incorporated adversarial training to further improve the robustness of our solution. Experiments on two datasets, namely UNSW-NB15 and NSL-KDD, demonstrate that the proposed approach increases resilience to adversarial attacks.
\end{abstract}

%%%%%%%%%%%%%%%%%%%%%%%%%%%%%%%%%%%%%%%%%%%%%%%%%%%%%%%%%%%%%%%%%%%%%%%%%%%%%%%%
\section{Introduction}
\begin{figure}
    \centering
    \includegraphics[width=0.9\linewidth]{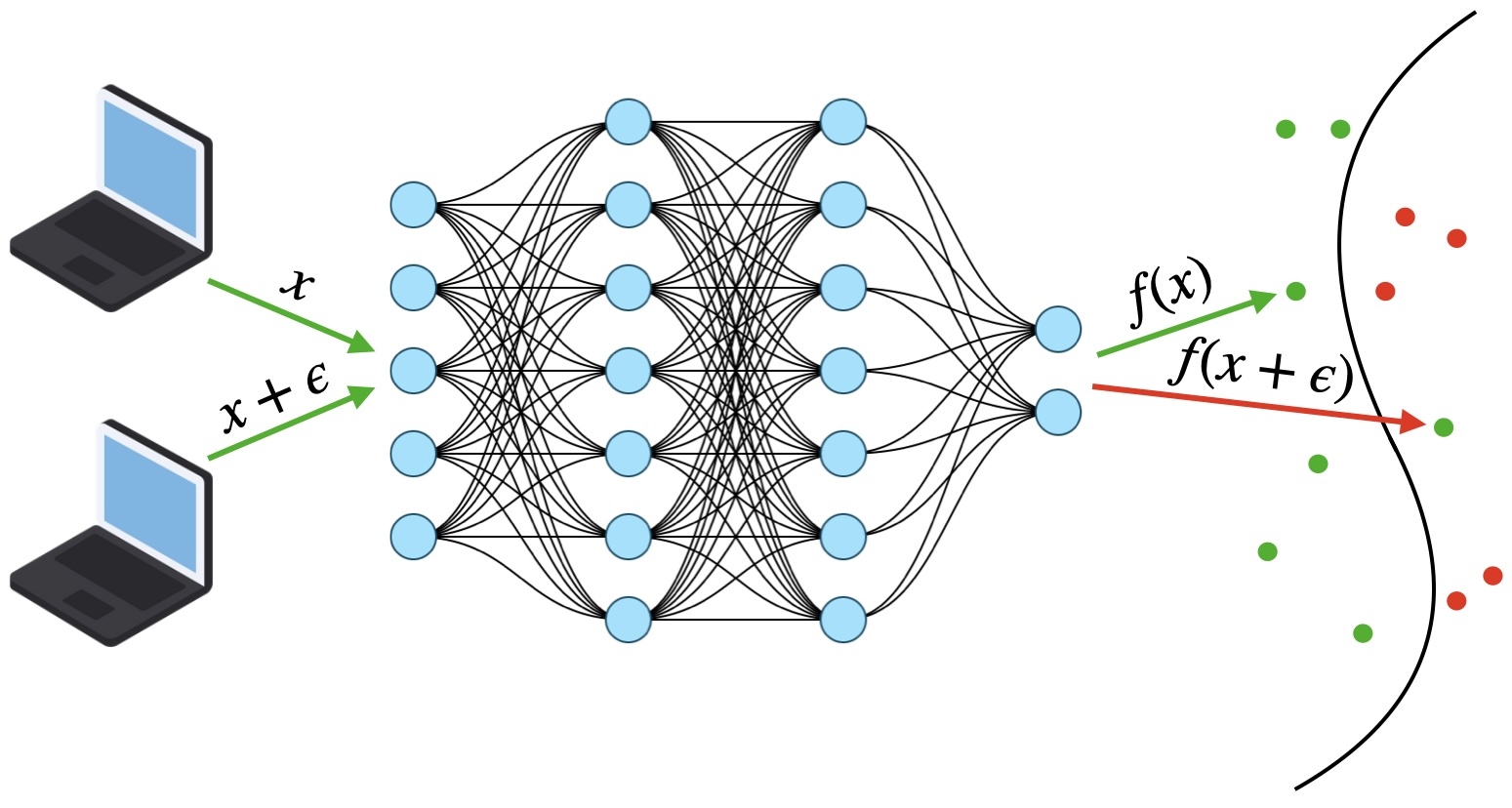}
    \caption{Under a small perturbation, $\epsilon$, input $x$ becomes an adversarial example that is misclassified into $f(x+\epsilon)$, which falls on the wrong side of the decision boundary.}
    \label{fig:aml}
\end{figure}

A Machine Learning (ML) model might incorrectly classify an input if it has been carefully crafted for this purpose as part of an adversarial attack \cite{kurakin2016adversarial}. This is typically achieved by adding a small perturbation $\epsilon$ to the input. This perturbation, often imperceptible, is enough to alter the model's behaviour, such that $f(x) \neq f(x+\epsilon)$. As shown in Figure. \ref{fig:aml}, the adversarial example is designed in such a way as to make the ML model map the input to the wrong class. The consequences of successful attacks can be far-reaching, including  confidentiality and privacy breaches, or  violations of integrity and availability, impacting the reliability of ML systems in critical applications such as medical diagnosis~\cite{finlayson2019adversarial} and Network Intrusion Detection Systems (NIDS) \cite{alatwi2021adversarial}.
To address this class of vulnerabilities, Adversarial Machine Learning (AML) emerges as a field that evaluates ML models against malicious attacks and proposes countermeasures to mitigate their detrimental effects \cite{he2023adversarial}. %This is particularly important to improve the resilience of computer network security.

NIDS monitors network traffic to detect malicious behavior \cite{ahmad2021network}. As a defense mechanism, its resilience to adversarial attacks is crucial, and this problem has been addressed in the recent literature. For example, Alhajjar et al.   \cite{alhajjar2021adversarial},  focus on measuring the vulnerability of NIDS to adversarial attacks capable of evading ML-based models. The adversarial inputs are generated using three techniques: particle swarm optimization (PSO), Genetic Algorithm (GA), and Generative Adversarial Network (GAN). The authors were able to increase the misclassification rates across several ML models, highlighting the need for more resilient NIDS. Similarly, Han et al. \cite{han2021evaluating},  evaluate the robustness of ML-based NIDS against traffic-space adversarial attacks. The generated examples were based on gray- and black-box methods and achieved over 97\% evasion. The authors also proposed a defense mechanism that reduced evasion to 50\%. The proposed attacks and the robustness of the defense method were evaluated on the state-of-the-art NIDS Kitsune and on several ML-based NIDS.
Abou Khamis et al.  \cite{abou2020investigating},  propose an intrusion detection system based on a min-max approach. Adversarial attack samples were generated using the UNSW-NB15 network dataset. Their defense mechanism relied on optimizing a model for intrusion detection, and the authors report that their approach was able to increase the robustness of NIDS. 

To remain effective, NIDS must withstand adversarial attacks while continuously integrating new anti-manipulation strategies \cite{ahmad2024data}. Hence, the aim of this work is twofold. First, we seek to create adversarial attacks to assess the vulnerabilities of traditional ML classifiers. Two network communication datasets, namely UNSW-NB15 \cite{dataset_unsw} and NSL-KDD \cite{nsl_kdd}, are used to generate network adversarial examples. We also propose a multi-layer solution that integrates a stacking classifier and an autoencoder to detect malicious network traffic. The contributions of this paper are as follows:

\begin{itemize}
    \item We evaluate the impact of two adversarial methods to generate malicious network traffic on the effectiveness of  ML-based NIDS. %These methods are fundamental to evaluate the resilience of ML-based NIDS towards adversarial attacks.
    \item We propose a defense mechanism based on a multi-layer approach and adversarial training, which successfully enhances the robustness of NIDS in two benchmarks.    
\end{itemize}

The remainder of this paper is structured as follows: Section \ref{Methodology} presents the mechanism used to generate adversarial attacks. Section \ref{defense} describes the proposed defense mechanism. Section \ref{Experimental Setup} details the experimental setup and Section \ref{Experimental Results} reports our results. Section \ref{Conclusion and future work } concludes the paper.

\section{Attack Methodology}
\label{Methodology}

\begin{figure}%[htbp]
\centering
    \includegraphics[width=.99\linewidth]{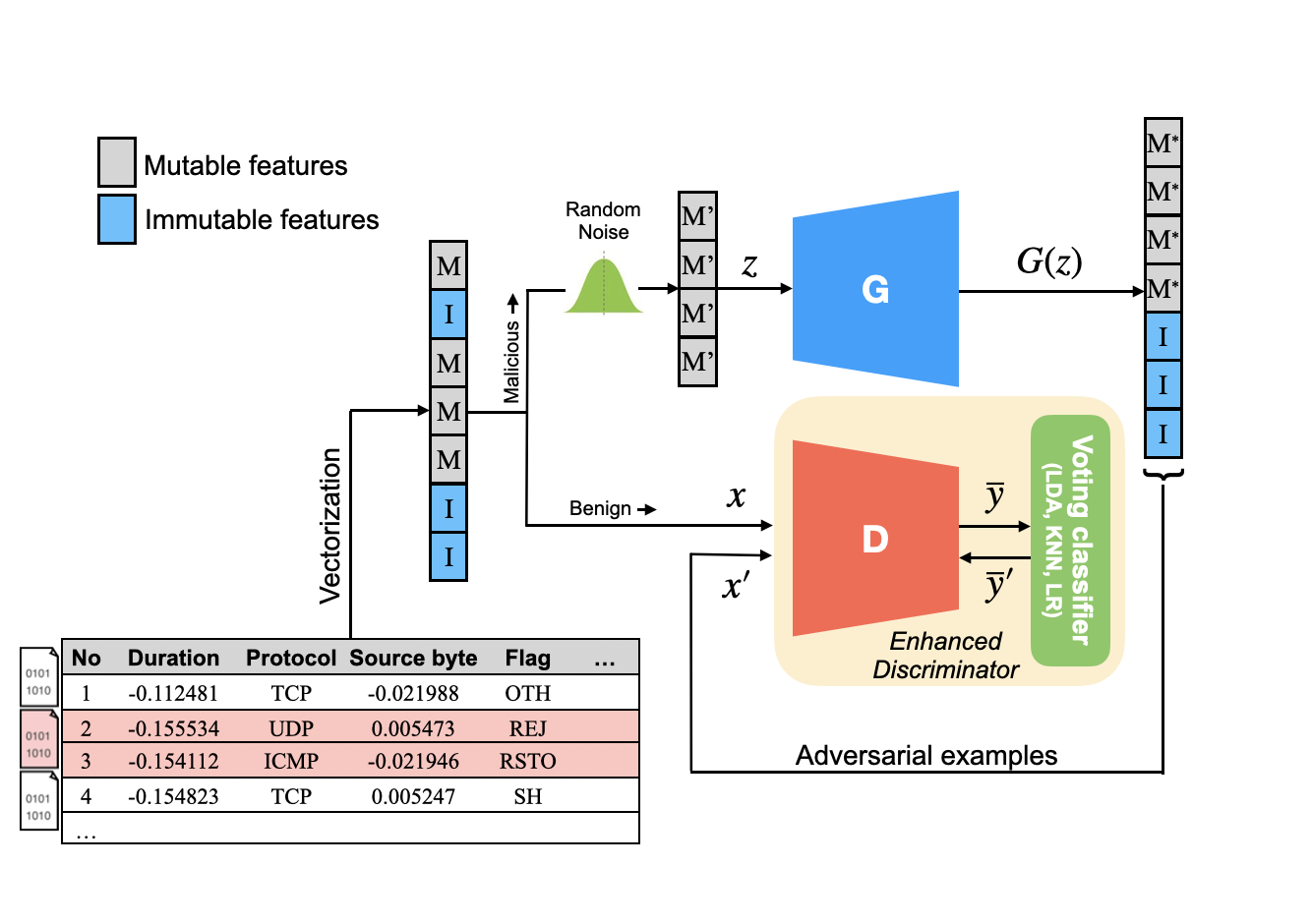}
    \caption{The Generative Adversarial Network against NIDS is composed of a generator $G$, a discriminator $D$ and a voting classifier. While $G$ is trained to produce mutable adversarial vectors, $D$ is optimized to identify malicious generated network traffic in collaboration with the voting classifier. }
    \label{fig:GAN}
\end{figure}

%In this section, we describe the two network adversarial attacks strategies employed in this paper. The first method is based on Generative Adversarial Networks (GAN) and the second one is the Fast Gradient Sign Method (FGSM). 
In this section, we describe the two network adversarial attack strategies used in this study: Generative Adversarial Networks (GAN) and the Fast Gradient Sign Method (FGSM), where FGSM serves as a baseline for comparison due to its simplicity and efficiency.

%\subsection{Problem Formulation}

\subsection{Generative Adversarial Network}
Inspired previous research  \cite{alhajjar2021adversarial,duy2021digfupas}, we propose the GAN architecture depicted in Figure 2. Note that after vectorizing the network traffic into mutable features $M$, and immutable features $I$, the data is separated into malicious and benign traffic. For the datasets adopted in this work, we considered the same immutable and mutable features defined by Alhajjar et al.  \cite{alhajjar2021adversarial}. The generator $G$, receives malicious input $z$,  which is mutable features perturbed by noise, and referred to as $M'$. Note that the immutable features are not used as input in this case. The generator outputs $M^*$, which is concatenated with $I$, producing new adversarial examples. The discriminator, $D$, receives as input real benign samples from the original network traffic as well as crafted malicious (i.e., adversarial examples) ones from the generator $G$. This approach ensures that malicious and crafted examples are produced by the generator while real and benign ones come from the original datasets. 

Similarly to the min-max approach presented by Kamis et al.  \cite{abou2020investigating}, the generator is trained to deceive the discriminator with malicious examples and the discriminator is optimized to detect adversarial traffic. This competition helps to improve both networks until the generator produces realistic outputs that fool the discriminator into misclassifying malicious samples as benign. In our work, we add an additional classifier, namely the \textit{voting classifier} , which we shall discuss next. Equation 1 shows the loss function for both networks, where $N$ is the total number of samples, $y_i$ represents the actual label (i.e., 1 for malicious, 0 for benign), and $p(y_i)$ is the predicted probability of maliciousness. 

%\begin{equation}
%\min_{G}\max_{D}V(G,D)=\mathbb{E}_{x\sim p_{\text{data}}(x)}[\log{D(x)}] +  \mathbb{E}_{z\sim p_{\text{z}}(z)}[1 - \log{D(G(z))}]
%\end{equation}

%\subsection{Voting Classifier}

A voting classifier is typically used to enhance the discriminator's performance \cite{khan2022voting}. In our study, we opted for three ML-based methods, namely k-Nearest Neighbors (KNN),  Linear Discriminant Analysis (LDA) and  Logistic Regression (LR) algorithms. We selected these voting classifier due to their complementary strengths, KNN captures local patterns, LDA handles linear separability with low variance, and LR offers probabilistic reasoning with regularization. This diversity helps improve robustness and reduce model bias.

These classifiers are referred to as $C_i$, with $i \in \{1,2,3\}$. The sample, $x$, is shared with the voting classifier, along with the discriminator output, $y'$. This information is used to create the final prediction, as described below:

\begin{equation}
    y' = majority-vote(C_1(x), C_2(x), C_3(x), \overline{y})
\end{equation}

Thus, after the sample is classified as malicious or benign by the voting classifier, the discriminator will then adjust its weights accordingly. This ensemble approach enhances the discriminator classification's accuracy and helps to further challenge the generator into creating more realistic adversarial samples. Note that the voting classifier is trained offline and then used during the optimization of the adversarial neural network. 

\begin{figure*}
\centering
    \includegraphics[width=0.6\linewidth]{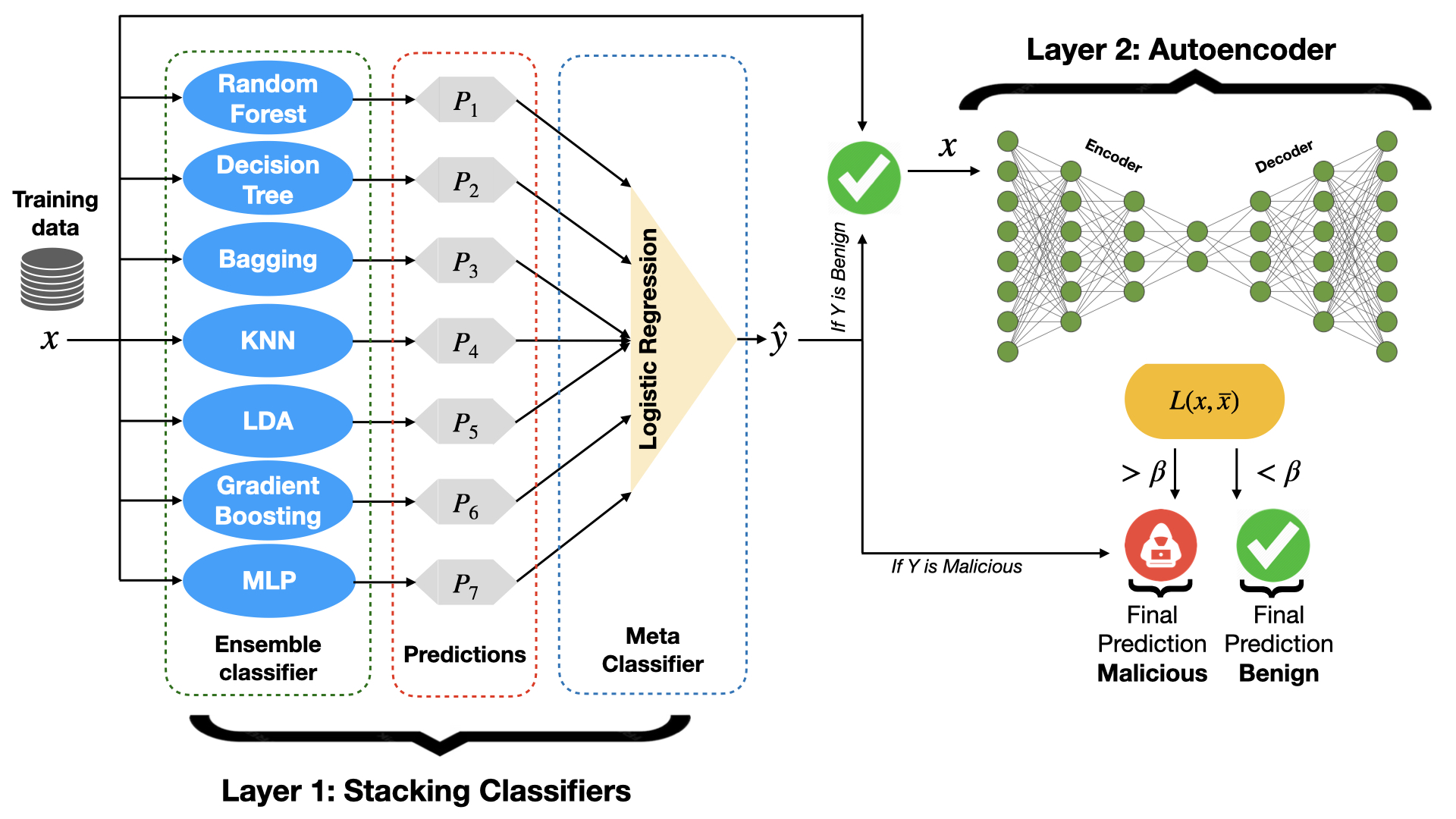}
    \caption{Proposed multi-layer solution for detecting adversarial network attacks.}
\label{solution}
\end{figure*}

\subsection{Fast Gradient Sign Method (FGSM)}
Typically, a neural network reduces its loss by adjusting the weights using the gradients obtained during back-propagation. The FGSM, on the other hand, generates adversarial examples by perturbing the gradients of a neural network\cite{goodfellow2014explaining}. The mathematical formulation for the FGSM attacks is given  by Equation~\ref{eq:fgsm}:

\begin{equation}
    x' = x + \epsilon \cdot \text{sign}(\nabla_x J(\theta, x, y))
    \label{eq:fgsm}
\end{equation}

%Here, \( x \) is the input, \( \epsilon \) scales the perturbation applied to mutable features, and \( J(\theta, x, y) \) is the loss gradient with respect to \( x \). The crafted perturbations generate adversarial malicious traffic aimed at deceiving a three-layer MLP classifier distinguishing between normal and anomalous network data\cite{haroon2022adversarial}.%

Here, \( x \) represents the input data, \( \epsilon \) defines the strength of the perturbation for mutable features as a scaling value, and \( J(\theta, x, y) \) denotes the gradient of the loss function with respect to the input. In this study, the gradients of the loss function, with respect to the input traffic's mutable features, are used to craft a modified traffic sample that maximizes the loss, leading to network adversarial malicious examples \cite{haroon2022adversarial}. A Multi-layer Perceptron (MLP) with three fully connected layers is used to classify network traffic as normal or anomalous, serving as the target model for evaluating the effectiveness of the generated adversarial attacks.

%Figure \ref{fig2} illustrates network traffic including malicious and benign samples based on samples from the original datasets, as well as generated adversarial examples. %%The samples are represented, respectively, in red, green and blue. The plots show some level of clusterization between the three classes, specially for the NSL-KDD dataset. In the UNSW-NB15, we can observe more malicious generated samples overlapping with the malicious samples from the original dataset. Next, we discuss how this affects the performance of our models.

\section{Defense Mechanism}
\label{defense}

Our defense mechanism is based on a two-layer solution comprised of a stacking classifier, an autoencoder layer and an adversarial training. We combine stacking, autoencoders, and adversarial training to enhance both accuracy and robustness. Stacking improves generalization, the autoencoder detects missed anomalies, and adversarial training strengthens resistance to crafted attacks. Details for each layers are provided below. 

\subsection{Stacking Classification} The first layer in our solution consists of an ensemble of ML classifiers. We adopted  Random Forest (RF), Decision Tree (DT), Bagging, K-Nearest Neighbors (KNN), Linear Discriminant Analysis (LDA), Gradient Boosting (GB) and MLP. This approach has been shown to effectively reduce the rate of misclassification \cite{rokach2010pattern}, which motivated its use in our study. The predictions of the ensemble classifier are aggregated and serve as the input of the meta classifier, represented here by a Logistic Regression (LR) model. The stacking classifier reduces the dangers of relying solely on one model, which may have biases and flaws. If the input sample $x$ is flagged as malicious, it is immediately classified as such and the final prediction is reached ( Figure \ref{solution}). However, if $x$ is classified as benign by the first layer, the input is sent to the autoencoder for further analysis.  This two-tiered defense setup leverages the broad decision-making strengths of ensemble classifiers alongside the anomaly detection sensitivity of autoencoders, enhancing the overall effectiveness of the intrusion detection system.

% This approach aims at minimizing the number of false negatives. The stacking is provided for diverse perspectives on network traffic, reducing variance, and mitigating overfitting. This ensures that if one classifier is deceived by an adversarial example, others may still correctly classify it, leading to a more reliable decision-making process.

\subsection{Autoencoder}

\begin{table*}
\centering
\caption{Performance on the NSL-KDD dataset for unmodified, GAN and FGSM conditions (p-value $<\;$ 0.01).}
\resizebox{0.99\textwidth}{!}{
\begin{tabular}{lccccccccccccccc}
\hline
& \multicolumn{3}{c}{Unmodified} & & \multicolumn{3}{c}{GAN} & & \multicolumn{3}{c}{FGSM} & & \multicolumn{3}{c}{All} \\
\cline{2-4} \cline{6-8} \cline{10-12} \cline{14-16}
& Precision & Recall & F1 & & Precision & Recall & F1 & & Precision & Recall & F1 & & Precision & Recall & F1 \\
\hline
DT   & 80.85 & 78.02 & 78.04 & & 75.24 & 62.77 & 64.10 & & 83.87 & 77.84 & 78.76 & & 80.28 & 66.05 & 68.50 \\
GB   & 80.05 & 80.06 & 80.05 & & 77.05 & 72.11 & 73.22 & & 84.47 & 84.13 & 84.27 & & 81.50 & 76.60 & 77.95 \\
BGG  & 81.51 & 78.83 & 78.86 & & 79.50 & 66.23 & 67.39 & & 83.91 & 76.97 & 77.95 & & 83.30 & 68.48 & 70.75 \\
KNN  & 83.55 & 77.35 & 77.05 & & \textbf{85.29} & 75.28 & 76.30 & & 88.21 & 82.88 & 83.58 & & \textbf{88.22} & 79.88 & 81.29 \\
LR   & 80.71 & 76.75 & 76.67 & & 77.29 & 58.45 & 59.01 & & 80.92 & 69.22 & 70.40 & & 80.60 & 58.11 & 60.40 \\
MLP  & 84.27 & 79.07 & 78.90 & & 83.62 & 70.43 & 71.48 & & 86.07 & 77.85 & 78.78 & & 85.90 & 71.89 & 73.93 \\
LDA  & 82.89 & 77.99 & 77.84 & & 78.97 & 57.19 & 57.25 & & 83.74 & 73.17 & 74.25 & & 82.52 & 59.47 & 61.66 \\
Ours & \textbf{84.64} & \textbf{84.42} & \textbf{84.23} & & 77.16 & \textbf{78.11} & \textbf{77.24} & & \textbf{89.37} & \textbf{89.47} & \textbf{89.22} & & {87.81} & \textbf{86.51} & \textbf{84.64} \\
\hline
\end{tabular}
}
\label{tab:nsl_full_performance}
\end{table*}

The autoencoder used in the second layer of our proposed NIDS is meant to refine the detection process by leveraging autoencoder ability to learn representations of benign traffic. To that end, the autoencoder is trained offline to reconstruct the input features. The model is optimized exclusively on benign traffic to reconstruct normal patterns with minimal reconstruction error \cite{sadaf2020intrusion}. The reconstruction loss adopted is the Mean Squared Error (MSE) between the original input $x$ and its reconstructed counterpart $\overline{x}$, as shown in Equation~\ref{eq:reconstruction_error}:

\begin{equation}
\label{eq:reconstruction_error}
L(x, \overline{x}) = \frac{1}{n} \sum_{i=1}^n (x_i - \overline{x}_i)^2
\end{equation}

\noindent where $n$ is the number of features. The architecture consists of a feedforward neural network with four fully connected layers: an input layer matching the feature dimension, two hidden layers of sizes 64 and 32 (forming a bottleneck), followed by a symmetric decoder that reconstructs the original input. If the error exceeds a threshold $\beta$, the traffic is flagged as malicious. This approach effectively filters out malicious traffic that bypasses the first layer, providing an additional safeguard for identifying sophisticated attacks while minimizing false negatives.  We empirically determine the anomaly detection threshold $\beta$ by selecting the 95$^{th}$ percentile of reconstruction errors on the validation set. Any sample whose reconstruction error exceeds this threshold is reclassified as malicious.

\subsection{Adversarial Training} Adversarial samples generated by the GAN and FGSM described in Section \ref{Methodology} are incorporated into the training set. Exposing the ML model to a portion of perturbed attack patterns during training phase helps  enhance the model’s robustness. Consequently, adversarial traffic flows are generated and used to train our stacking classifier. By integrating these perturbed samples into the original training data, the model learns to recognize and counter adversarial variations. To assess generalization, adversarial samples from one method (e.g., FGSM) are used to test a model being training with another method, such as  GAN. This cross-testing improves the stacking classifier's ability to detect unseen adversarial patterns.

\section{Experimental Setup}
\label{Experimental Setup}
\subsection{Datasets}
\subsubsection{NSL-KDD Dataset}
The NSL-KDD dataset provides network traffic data to train and test NIDS. There are 22,544 network traffic records in the test set and 125,973 records in the training set. Each record consists of 41 attributes, which include basic features of each network connection vector (such as duration, protocol type), content-related features (such as hot, is\_guest\_login), and traffic features based on two-second time windows (such as count, srv\_count, serror\_rate).  This dataset aids in the meticulous analysis of network interactions, facilitating the detection of anomalies and intrusions \cite{dhanabal2015study,soltani2024robust}. 
%Attack types in the NSL-KDD dataset are categorized into four main groups: DoS (Denial of Service), Probe (Surveillance/Probing), U2R (Unauthorized Access to Local Super User Privileges), and R2L (Unauthorized Access from a Remote Machine).

\subsubsection{UNSW-NB15}
The UNSW-NB15 dataset, crafted by the Australian Centre for Cyber Security (ACCS), serves as a critical benchmark for testing the efficacy of NIDS. The dataset comprises 257,673 entries, with 175,341 dedicated to training and 82,332 allocated for testing purposes. It spans 49 distinct features per entry, incorporating an extensive range of network flow attributes and derived metrics designed to encapsulate the intricate dynamics of network traffic. The attack categories featured include Fuzzers, Analysis, Backdoors, DoS, Exploits, Generic, Reconnaissance, Shellcode, and Worms, each contributing to the dataset’s diversity and complexity.

\subsection{Data Preprocessing}
The UNSW-NB15 and NSL-KDD datasets contain features of various types, including both numerical and categorical values requiring comprehensive pre-processing to ensure compatibility with ML models. The pre-processing pipeline consists of two stages: data transformation and value normalization \cite{duy2021digfupas}. To handle the heterogeneous nature of the datasets, the features are divided into numerical, categorical, and label columns. Missing numerical attributes are filled with median values. Categorical features like service and flag are first imputed with their most frequent value when missing, and then converted into numerical form using ordinal encoding. Normalization is applied to numerical features, to mitigate the effects of varying scales between features.  For instance, features that represent counts may have values within a small range, while features such as duration may span several orders of magnitude. We employ standard scaling  to normalize numerical features \cite{duy2021digfupas}, transforming them to have a mean of zero and a standard deviation of one. This ensures improved convergence during training and robust performance during testing.

\begin{table*}
\centering
\caption{Performance on the UNSW-NB15 dataset for unmodified, GAN and FGSM conditions (p-value $<\;$ 0.01).}
\resizebox{0.95\textwidth}{!}{
\begin{tabular}{lccccccccccccccc}
\hline
& \multicolumn{3}{c}{Unmodified} & & \multicolumn{3}{c}{GAN} & & \multicolumn{3}{c}{FGSM} & & \multicolumn{3}{c}{All} \\
\cline{2-4} \cline{6-8} \cline{10-12} \cline{14-16}
& Precision & Recall & F1 & & Precision & Recall & F1 & & Precision & Recall & F1 & & Precision & Recall & F1 \\
\hline
DT   & 61.18 & 58.32 & 57.69 & & 65.87 & 62.87 & 63.18 & & 67.90 & 64.88 & 65.18 & & 62.98 & 55.29 & 54.89 \\
GB   & 76.52 & 76.15 & 75.75 & & 74.33 & 74.58 & 74.36 & & 75.03 & 75.28 & 75.02 & & 74.43 & 74.68 & 74.51 \\
BGG  & 59.26 & 56.66 & 56.05 & & 60.01 & 54.52 & 54.18 & & 61.81 & 57.25 & 57.29 & & 69.73 & 68.02 & 68.42 \\
KNN  & 82.99 & \textbf{81.88} & \textbf{81.47} & & 82.69 & \textbf{82.40} & \textbf{81.99} & & 84.03 & \textbf{83.47} & \textbf{83.01} & & 83.53 & \textbf{83.30} & \textbf{82.86} \\
LR   & 83.40 & 77.76 & 76.12 & & 73.73 & 73.55 & 72.38 & & 73.96 & 73.73 & 72.54 & & 71.72 & 72.23 & 71.28 \\
MLP  & {84.25} & 81.71 & 81.02 & & 79.23 & 79.09 & 78.56 & & 80.08 & 79.78 & 79.21 & & 78.51 & 78.67 & 78.20 \\
LDA  & 82.28 & 73.90 & 71.07 & & 71.32 & 70.69 & 68.52 & & 72.55 & 71.47 & 69.20 & & 69.71 & 70.12 & 68.17 \\
Ours & \textbf{85.04} & 73.95 & 73.62 & & \textbf{84.25} & 81.16 & 79.84 & & \textbf{84.81} & {80.00} & {78.14} & & \textbf{83.73} & 80.35 & 78.64 \\
\hline
\end{tabular}
}
\label{tab:unsw_full_performance}
\end{table*}

\section{Experimental Results}
\label{Experimental Results}
In this section, we present the results of the experimentation conducted to assess our model and baseline systems.

\subsection{Network Adversarial Detection}

Table \ref{tab:nsl_full_performance} presents the performance comparison between the proposed model—an ensemble learning approach enhanced with adversarial training—and baseline machine learning classifiers trained solely on original (non-adversarial) malicious and normal traffic. This comparison highlights both the vulnerability of traditional models to adversarial attacks and the robustness gains achieved by integrating adversarial training with ensemble learning. 

The performance of each model degraded in the presence of adversarial examples generated by the GAN method. The  DT, LR and LDA models were the most affected. The  \textit{F1-score} for these models fell to  64.10\%, 59.01\% and 57.25\% respectively,   a decrease of performance of 13.92\%, 17.66\% and 20.59\%, (resp.).  KNN is  the least affected by the GAN adversarial examples, with less than 1\% decay,  from 77.05\% to 76.30\%. The proposed two-layer classifier, on the other hand, provided the highest F1-score, achieving 84.23\%, 77.24\% and 89.22\%, respectively, for the unmodified, GAN and FGSM conditions. While the  GAN-based approach does incur  a significant decay of 6.99\%, it still outperforms KNN, which was the second best model in terms of F1-score. KNN provided the best \textit{Precision}, 85.29\%, followed by the MLP model, 83.62\%. In general, \textit{Recall} was the most affected by all ML baselines when facing the GAN-based adversarial networks, with our proposed method being the most robust one. Overall, the models were not as strongly impacted by the FGSM attacks, with the proposed two-layer solution reaching the highest performance. In the presence of all attacks, the proposed approach achieved the highest \textit{Recall} and \textit{F1-score}.

\begin{figure*}[t]
    \centering
    \includegraphics[width=0.90\textwidth]{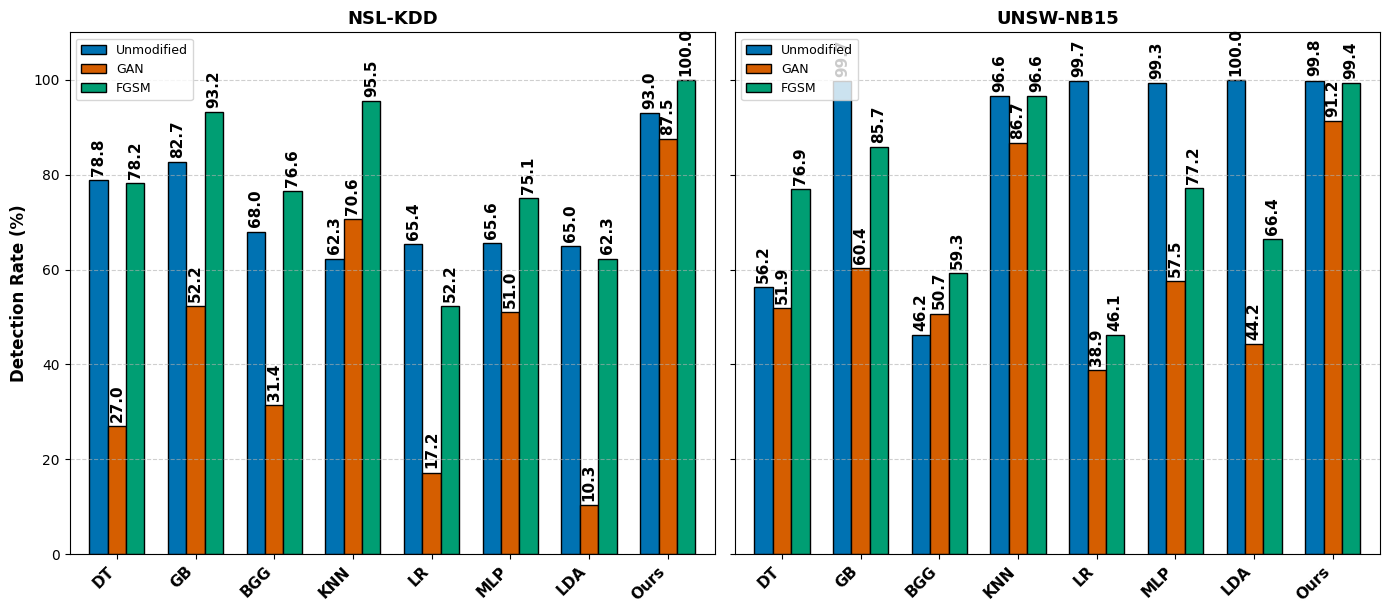}
    \caption{Detection rate comparison of ML classifiers on the NSL-KDD and UNSW-NB15 datasets under unmodified, GAN and FGSM conditions.}
    \label{fig:detection_rate_comparison}
\end{figure*}

In Table \ref{tab:unsw_full_performance}, we present the results for the UNSW-NB15 dataset. Compared to the previous dataset, the results are slightly lower across all conditions, since UNSW-NB15 includes a wider variety of attack types, deceiving will be more challenging. DT is still significantly impacted by the adversarial attacks, experiencing a decay of 11.58\% and 9.03\%, for the GAN and FGSM conditions respectively. The LR was the second most impacted model, decaying by 4.84\% and 4.21\% (resp.) for the GAN and FGSM approaches. KNN provides the best \textit{F1-score} for all conditions, followed by our proposed solution for adversarial attacks based on the GAN and FGSM methods. Our solution yields the best precision in all scenarios, revealing that the model detects the attacks with high accuracy, but misses few true positives as the \textit{Recall} is around 9\% lower than \textit{Precision}. These results are due to the second layer (i.e., the autoencoder) which assumes that the negative predictions made by the stacking classifier may be false negatives and must be further verified. Note that results in terms of detection rate are provided in Figure \ref{fig:detection_rate_comparison}. We found that the proposed two-layer solution provides the best results. Results show that models such as  DT, BGG, and LDA experience significant decay under adversarial attacks, especially GAN. In contrast, GB, KNN, and MLP demonstrate better resilience, with KNN maintaining performance under FGSM attacks. Shown in Figure \ref{fig:detection_rate_comparison}, the proposed model achieves the highest detection rates under GAN attacks, reaching an DR of 87.55\% on the NSL-KDD 91.24\% on the UNSW-NB15 datasets, while and under FGSM attacks it achieves 100\% on the NSL-KDD dataset, which highlights the effectiveness of the proposed defense mechanism with adversarial training. We also compare the accuracy of the proposed model with that of deep neural network models. These results, shown in Table \ref{tab:comparison_accuracy}, are based on the original and unmodified datasets, not including any adversarially generated samples. The proposed model outperforms all the baselines considered.

\begin{table}
\centering
\caption{Comparison with deep neural network models (accuracy).}
\scalebox{1.15}{
\begin{tabular}{ccc}
\hline
Dataset & Models                  & Acc \\ \hline
\\{NSL-KDD} & BAT-MC \cite{su2020bat}          & 84.15                   \\ %\cline{2-3}
                         & Autoencoder \cite{sanaboina2023examining}     & 84.21                   \\ %\cline{2-3}
                         %& Imamverdiyev et al. \cite{ribeiro2020lime} From  \cite{sanaboina2023examining} & 73.23                  \\ %\cline{2-3}
                         & Ours           & \textbf{87.35}                 \\  \\ \hline
\\{UNSW-NB15} & Scaling SVM \cite{jing2019svm}    & 75.77                   \\ %\cline{2-3}
                           & DT (C5) \cite{meftah2019network}        & 75.53                   \\ %\cline{2-3}
                           & Ours          & \textbf{77.71}              \\   \\ \hline
\end{tabular}}
\label{tab:comparison_accuracy}
\end{table}

\subsection{Impact of adversarial training}
Table \ref{tab:adversarial_training_performance} presents an ablation study evaluating the effect of adversarial training and the integration of an autoencoder to our stacking solution. We evaluate the two datasets and observe that the stacking classifier (SC) benefits from  the inclusion of the autoencoder, specially for the NSL-KDD dataset. Notably, adversarial training significantly boosts detection rates, with SC+AE achieving 92.99\% on the NSL-KDD dataset and 99.97\% on the UNSW-NB15 dataset. These results show that combining adversarial training with anomaly detection leads to more robust and effective   detection of adversarial attacks.

\begin{table}
\centering
\renewcommand{\arraystretch}{1.5} % Increase row height
\caption{Ablation study considering original data (OD), adversarial (AT) training, the stacking classifier (SC) and the latter with anomaly detection including the autoencoder (AE).}
\scalebox{1.15}{
\begin{tabular}{cccccc}
\hline
Dataset                                & OD         & AT & SC & AE & DR \\ \hline
{NSL-KDD}               & \cmark & \xmark & \cmark & \xmark   & 72.42                   \\ %\cline{3-4}
                                       & \cmark & \xmark & \cmark & \cmark  & 77.80                   \\ %\cline{2-4} 
                                                 & \cmark & \cmark & \cmark & \xmark       & 90.29                   \\ %\cline{3-4}
                                                 & \cmark & \cmark & \cmark & \cmark  & \textbf{92.99}                   \\ \hline
{UNSW-NB15}            & \cmark & \xmark & \cmark & \xmark  & 86.05                   \\ %\cline{3-4}
                                                 & \cmark & \xmark & \cmark & \cmark  & 86.23                   \\ %\cline{2-4}
                                                 & \cmark & \cmark & \cmark & \xmark & 99.78                  \\ %\cline{3-4}
                                                & \cmark & \cmark & \cmark & \cmark & \textbf{99.97}                   \\ \hline
\end{tabular}}
\label{tab:adversarial_training_performance}
\end{table}

\section{Conclusion}
\label{Conclusion and future work }
In this study, we tested the effectiveness of NIDS against two classes of adversarial attacks, namely GAN and FGSM, using augmented NSL-KDD and UNSW-NB15 datasets. Our study reveals critical insights into the robustness of various ML classifiers operating in this context. The presence of advanced adversarial tactics underscores the need for evolving NIDS methodologies that can counteract these threats. Our proposed hybrid classifier adopts a two-layer design that combines stacking and autoencoder models. After applying adversarial training, it achieves a detection rate of approximately 92\% on the NSL-KDD dataset and 99\% on the UNSW-NB15 dataset under unmodified conditions. The detection rate also improves against adversarial traffic, reaching around 99\% for FGSM-based attacks and approximately 90\% for GAN-generated samples. To address evolving adversarial threats, adversarial machine learning should continue improving hybrid and ensemble approaches. Ensuring the robustness of NIDS requires regular retraining with emerging attack strategies. Additionally, improving model accuracy and robust feature engineering, along with enhancing computational efficiency, is essential for scalable deployment in multi-class intrusion scenarios.

\bibliographystyle{IEEEtran}
\bibliography{bib}

\end{document}